\documentclass[aps, twocolumn, superscriptaddress]{revtex4-2} 

\usepackage[switch]{lineno}
\usepackage{amsmath}
\usepackage{amssymb}
\usepackage{empheq}
\usepackage[parfill]{parskip}
\usepackage[active]{srcltx}
\usepackage{color}
\usepackage{array}
\usepackage{booktabs}
\usepackage{amsfonts}
\usepackage{dsfont}
\usepackage{graphicx}
\usepackage{appendix}
\usepackage{hyperref}
\usepackage{lineno}
\usepackage{enumerate}

\begin{document}
\setlength{\parindent}{0.5cm}
\title{Amplitude responses of swarmalators}
\author{Samali Ghosh}
\altaffiliation{These authors contributed equally to this work}
\affiliation{Physics and Applied Mathematics Unit, Indian Statistical Institute, 203 B. T. Road, Kolkata 700108, India}

\author{Suvam Pal}
\altaffiliation{These authors contributed equally to this work}
\affiliation{Physics and Applied Mathematics Unit, Indian Statistical Institute, 203 B. T. Road, Kolkata 700108, India}

\author{Gourab Kumar Sar}
\altaffiliation{These authors contributed equally to this work}
\affiliation{Physics and Applied Mathematics Unit, Indian Statistical Institute, 203 B. T. Road, Kolkata 700108, India}

\author{Dibakar Ghosh}
\email{Corresponding to: diba.ghosh@gmail.com}
\affiliation{Physics and Applied Mathematics Unit, Indian Statistical Institute, 203 B. T. Road, Kolkata 700108, India}

\begin{abstract}
\hspace{1 cm}  (Received XX MONTH XX; accepted XX MONTH XX; published XX MONTH XX) \\ 

Swarmalators are entities that swarm through space and sync in time and are potentially considered to replicate the complex dynamics of many real-world systems. So far, the internal dynamics of swarmalators have been taken as a phase oscillator inspired by the Kuramoto model. Here, for the first time, we examine the internal dynamics utilizing an amplitude oscillator capable of exhibiting periodic and chaotic behaviors. To incorporate the dual interplay between spatial and internal dynamics, we propose a general model that keeps the properties of swarmalators intact. This adaptation calls for a detailed study which we present in this paper. We establish our study with the R{\"o}ssler oscillator by taking parameters from both the chaotic and periodic regions. While the periodic oscillator mimics most of the patterns in the previous phase oscillator model, the chaotic oscillator brings some new fascinating states.
\noindent 
\\

DOI: XXXXXX
\end{abstract}

\maketitle
\section{introduction}\label{section1}
Synchronization~\cite{pikovsky2001universal,winfree1980geometry,winfree1967biological} is one of nature's most fascinating and widespread phenomena. It refers to the coordination or alignment of events or processes in time, without necessarily influencing their spatial positions. This phenomenon is observed in a wide range of scales, from the microscopic to the macroscopic, and across various disciplines of science and nature. From cardiac pacemaker cells \cite{peskin1975mathematical}, the metabolic cycle of yeast cells \cite{aldridge1976cell}, coherently flashing fireflies \cite{buck1988synchronous,buck1938synchronous} to power grid dynamics \cite{motter2013spontaneous}, Josephson junction \cite{wiesenfeld1996synchronization} and even the unexpected wobbling of London's Millennium Bridge \cite{strogatz2005crowd}, there are so many examples that highlight the occurrence of spontaneous synchrony. Similarly, a complementary form of self-organization occurs in swarming \cite{couzin2002collective,couzin2007collective,sumpter2010collective,mogilner1999non} where individuals traverse through space, yet overtly modify their internal states. As evidence, birds fly in flocks \cite{bialek2012statistical,ballerini2008interaction}, fish swim in schools \cite{katz2011inferring}, and bacteria aggregates \cite{levy2008stochastic,chavy2016local} in space, yet do not always synchronize the timing of an internal state or rhythm. 
In a sense, both phenomena are spatio-temporal but opposites. Research on swarming such as the collective behavior of many body systems or self-propelled particles \cite{pal2024directional}, has garnered significant attention in recent decades. Studies of synchronization and swarming jointly establish a fruitful connection between physics and biology, delving into the influence of both spatial and temporal dependencies of the agents.

In recent decades, these two domains have been explored autonomously. Ultimately, the analysis of mobile oscillators bridged these two domains by exploring the impact of oscillator movement on their internal state \cite{frasca2008synchronization,majhi2019emergence,majhi2017synchronization}. However, the reverse scenario was not investigated. Mobile oscillators become the key element in modeling biological and robotic phenomena \cite{stilwell2006sufficient, frasca2008synchronization, fujiwara2011synchronization,buscarino2016interaction}. 
Von Brecht and Uminsky \cite{von2016anisotropic} proposed an aggregation model where the particles undergo an internal polarization vector comparable to the oscillator's phase. But in all these works, the relative distances between the particles affected their internal phase. Nevertheless, their relative phases did not significantly contribute to their actions. The theoretical exploration of the systems that feel the combined effects of swarming and synchronization started to garner interest about $17$ years ago. Tanaka \cite{tanaka2007general}, led the initiative by proposing a model of chemotactic oscillators where oscillators interact through a background diffusive chemical and produce diverse rich phenomena. Following this, recently O'Keeffe et al. \cite{o2017oscillators} introduced a toy model of swarming and synchronization, without reference to any background medium, called \textit{swarmalators} where the bidirectional interaction was considered, i.e., twin activities of the swarming oscillators can be tracked. They formulated this model by using space-dependent generalized globally coupled Kuramoto oscillators \cite{ott2008low} that anticipate several rich spatiotemporal patterns.

\par Since then there has been ongoing research to investigate further and understand the dynamics of swarmalators in different system-interaction configurations. The two-dimensional (2D) model was reduced to a one-dimensional (1D) solvable ring model \cite{o2018ring} which captures the behavior of swarm in quasi-1D rings such as sperms or vinegar eels \cite{peshkov2022synchronized}.
Using the Ott-Antonsen (OA) ansatz~\cite{10.1063/1.2930766}, the first analytical description of these states and the condition of their existence have become possible \cite{yoon2022sync}. To understand the influence of the surrounding environment, Hong et al.~\cite{hong2023swarmalators} introduced thermal noise in a population of swarmalators and came across several distinct collective patterns, some of which captivate the behavior of real-world swarmalators. In another work, the impact of the external damping force on the phase resulted in the phase transition from a nonforced model to full synchrony via partial synchronization \cite{lizarraga2020synchronization}. Several other aspects such as time-delayed interactions \cite{blum2024swarmalators}, distributed coupling \cite{o2022swarmalators}, finite cut-off interaction distance \cite{lee2021collective}, multiplexity \cite{kongni2023phase} and time-varying competitive phase interactions \cite{sar2022swarmalators} among the swarmalators give rise to a plethora of collective patterns. Sar et al. took the initiative to introduce random pinning \cite{sar2023pinning,sar2023swarmalators,PhysRevE.109.044603} subjected to the 1D ring model which delivers low-dimensional chaos and abrupt transition to a synchronous state along with phase wave and split phase wave states. Recently, swarmalators have been investigated under different community structures, and in the purely repulsive coupling, antiphase synchronization between the communities has been observed \cite{ghosh2023antiphase}. Check out the review articles on swarmalators in \cite{o2019review, sar2022dynamics}.

\par In most of the works done so far, the internal dynamics of swarmalators have been explored in terms of phase oscillators through the lens of the Kuramoto model. We envision swarmalators as a common playground of swarming and synchronization for which the oscillatory dynamics need not be controlled by the phase, and can also be rendered by the amplitude. In this work, we use one such amplitude oscillator model that governs the internal dynamics of the system. We choose parameters for the model such that the oscillations are chaotic. Synchronization of chaotic oscillators, where two or more chaotic systems evolve along similar chaotic trajectories despite starting from different initial conditions, represents a highly intriguing dynamical phenomenon that has been extensively investigated from various perspectives. In the context of swarmalators, this chaotic behavior may arise from the nonlinear interactions and feedback mechanisms among the individual components of the swarmalator system. This scenario is an apt depiction of challenges akin to the task coordination observed in swarming animals, which not only synchronize their movements within a two-dimensional plane but also respond collectively when subjected to threats or animal attacks. 
We study the ramifications that arise when the internal dynamics of the swarmalators are influenced by amplitude-mediated chaotic oscillators. We report our findings of some of the existing states of swarmalators for chaotic dynamics. In Sec.~\ref{section2}, we represent the fundamental mathematical model which serves as the basis for our study. We depict our results in Sec.~\ref{section3}. Lastly, we sum up with period one case and conclusions in Sec.~\ref{section4} and \ref{section5}, respectively.

\section{mathematical model}
\label{section2}
We consider $N$ number of swarmalators moving in a 2D region. The internal dynamics ($\mathbf{x}_i\in \mathbb{R}^d$) of the swarmalators are governed by a $d$-dimensional amplitude oscillator. The bidirectional interplay between spatial and internal dynamics of the swarmalators is given by the following set of equations,
\begin{align}
\Dot{\mathbf{r}}_i =&\mathbf{v}_i + \dfrac{1}{N}\sum_{j\neq i}^{N}\big(\mathbf{r}_j-\mathbf{r}_i\big)  \bigg[\big(A+J_1 e^{-E_{ij}}\big)-\dfrac{\big(B-J_2 e^{-E_{ij}}\big)}{|\mathbf{r}_j-\mathbf{r}_i|^{2}}\bigg],
\label{eq1}
\end{align}
\begin{align}
    \Dot{\mathbf{x}}_i = f(\mathbf{x}_i) + \dfrac{K}{N} \sum_{j\neq i}^{N} H (\mathbf{x}_i,\mathbf{x}_j,\mathbf{r}_i,\mathbf{r}_j),
    \label{eq2}
\end{align}
for $i=1,2,\ldots,N$ where $\mathbf{r}_i = (\xi_i,\eta_i) \in \mathbb{R}^2$ denotes the spatial position of the $i$-th swarmalator in the 2D plane. The attractive force among the swarmalators is represented by the 1st term inside the summation in Eq. \eqref{eq1} as $I_\text{att}= \big(A+J_1 e^{-E_{ij}}\big) $ and the second term defines the repulsive interaction among the swarmalators as $I_\text{rep}= \left(B-J_2 e^{-E_{ij}}\right) / |\mathbf{r}_j-\mathbf{r}_i |^{2}$. Here, $E_{ij}$ denotes the difference between internal dynamics of the $i$-th and the $j$-th swarmalators and is defined as $E_{ij} = |\mathbf{x}_j - \mathbf{x}_i|$, where $| \cdot |$ denotes the Euclidean norm. $A$ and $B$ are the strengths of spatial attraction and repulsion, respectively. $J_1$ and $J_2$ measure the influence of internal dynamics of the swarmalators on spatial attraction and repulsion, respectively.

$f: \mathbb{R}^d \rightarrow  \mathbb{R}^d$ in Eq.~\eqref{eq2} denotes the uncoupled identical internal dynamics of the swarmalators where $\mathbf{x}_i\in \mathbb{R}^d$. $H: \mathbb{R}^d \times \mathbb{R}^d \times \mathbb{R}^2 \times \mathbb{R}^2 \rightarrow  \mathbb{R}^d$ stands for the effect of spatial configurations on the internal dynamics. In our study, we employ the chaotic R{\"o}ssler attractor to describe the internal dynamics of the system, i.e.,
\begin{align}
\label{eq3}
f(\mathbf{x}_i)=
\begin{pmatrix}
    -y_i-z_i& \\
    x_i + a y_i&\\
    b + z_i(x_i-c)&
    \\
\end{pmatrix},
\end{align}
where $\mathbf{x}_i = (x_i,y_i,z_i) \in \mathbb{R}^3$. The interacting function $H$ defines the diffusive coupling between the $y$-components of the R{\"o}ssler oscillators, which is modulated by the spatial distance and is given by
\begin{equation}
\label{eq4}   H(\mathbf{x}_i,\mathbf{x}_j,\mathbf{r}_i,\mathbf{r}_j)=  \left[0, \frac{y_j-y_i}{|\mathbf{r}_j-\mathbf{r}_i|^\gamma}, 0\right ]^{\mathcal{T}},
\end{equation}
where $\mathcal{T}$ denotes the transpose of matrix and $K$ is the coupling strength. We are concerned only about the interaction through the $y$ components in our study. While dealing with the diffusive coupling via $x$ or $z$ components in the R{\"o}ssler oscillator, the Master Stability Function $(\Phi(K))$ \cite{wolf1985determining} concerning the coupling strength $(K)$ is always positive, which indicates that the coupled network of oscillators does not achieve any stable synchronization state. When considering interactions among the $y$ components in the R{\"o}ssler oscillator, $\Phi(K)$ is negative, i.e., the system concludes to achieve synchrony \cite{huang2009generic}.

In the rest parts of our investigation, we choose $N=200$ identical swarmalators with velocities $\mathbf{v}_i=0$ and fix $\gamma=1$.

\begin{figure*}
    \centering
    \includegraphics[width=2\columnwidth]{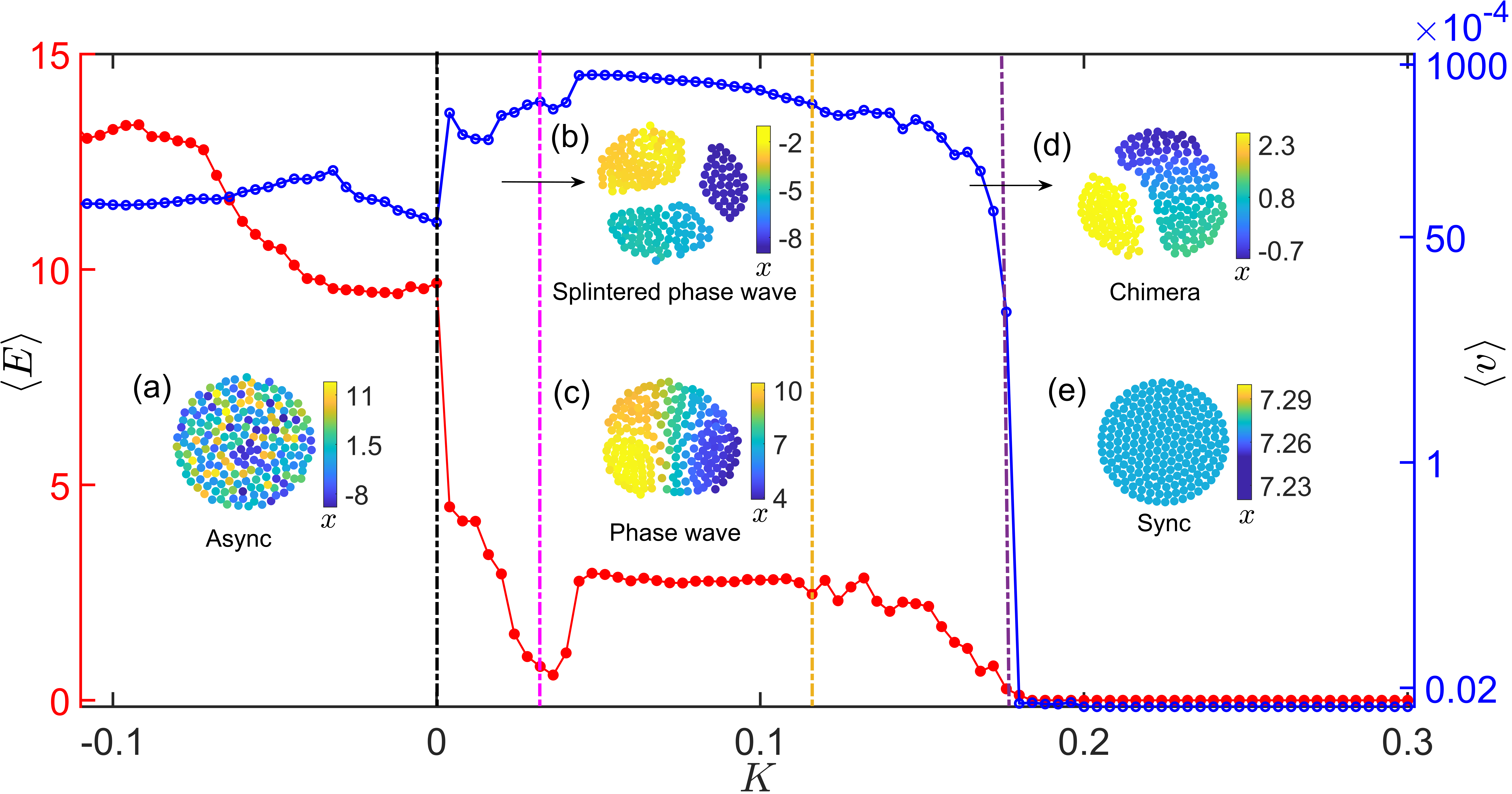}
	\caption{Measurement of the average synchronization error $\langle E \rangle$ of the R{\"o}ssler oscillator (filled red circles) and the mean velocity of the swarmalators $\langle v \rangle$ (open blue circles) as functions of the coupling strength $K$. The dotted vertical lines are used to distinguish distinct emerging states. (a) Async for $K=-0.068$, (b) splintered phase wave for $K=0.012$, (c) phase wave for $K=0.072$, (d) chimera for $K=0.132$, and (e) sync for $K=0.2$. We use the logarithmic scale to visualize $\langle v \rangle$ to differentiate the states by their velocity. The color bar signifies the $x$ component of the R{\"o}ssler oscillator. We fix $J_1=2.0$ and $J_2=0.5$. Heun's method has been used to simulate $N=200$ swarmalators with time-step $dt = 0.01$ for $T=4000$ time units. Initially, the swarmalators are placed uniformly randomly inside a square box of length $2$ centered at the origin. All the resultant states are observed after a long transient. The average synchronization error $\langle E \rangle$ and the mean velocity $\langle v \rangle$ are calculated after wiping out the first $75 \%$ data with $10$ realizations.}
	\label{fig1} 
\end{figure*}

\section{Results for chaotic system}
\label{section3} 

Our primary objective in this study is to explore the swarmalator field through the lens of chaotic oscillators. Hence we first focus on the chaotic region of the R{\"o}ssler oscillator. Later in Sec.~\ref{section4}, we also investigate the collective patterns in the case of period one R{\"o}ssler oscillator. Here we choose $a=b=0.2$, and $c=5.7$ belonging to the chaotic regime. In our study, every agent has three internal degrees of freedom ($x$, $y$, and $z$), and their internal dynamics arise from an ensemble of globally connected R{\"o}ssler oscillators. In the subsequent study, we choose $A=1$ and $B=2$. The other parameters $J_1$, $J_2$, and $K$ act as the primary control parameters. The spatial attraction term, $A+J_1 e^{-E_{ij}}$ increases as the relative difference between the internal dynamics of the swarmalators $(E_{ij})$ decreases. In contrast, the repulsion force among the swarmalators $(B-J_2 e^{-E_{ij}})$ diminishes, when the term $(E_{ij})$ decreases and vice versa. It can be readily comprehended in the following manner:

\begin{enumerate}[(i)]
\item 
When the difference of the internal dynamics among the swarmalators decreases, the term $e^{-E_{ij}}$ increases exponentially and finally attains the value unity. Hence, the attraction strength reduces to $A+J_1$ and that of repulsion becomes $B-J_2$. 
\item The opposite scenario occurs when $E_{ij}$ reaches a larger value so that the term $e^{-E_{ij}}\rightarrow 0$ and the strength of attraction and repulsion become independent of $J_1$ and $J_2$.
\end{enumerate}

Depending on the choice of these control parameters, the swarmalators exhibit various long-term emerging states, ranging from asynchronous to synchronous states. Before explaining these states, we define some order parameters that prove to be beneficial in quantifying various properties of the emerging states.

\begin{figure*}
    \centering
    \includegraphics[width=2\columnwidth]{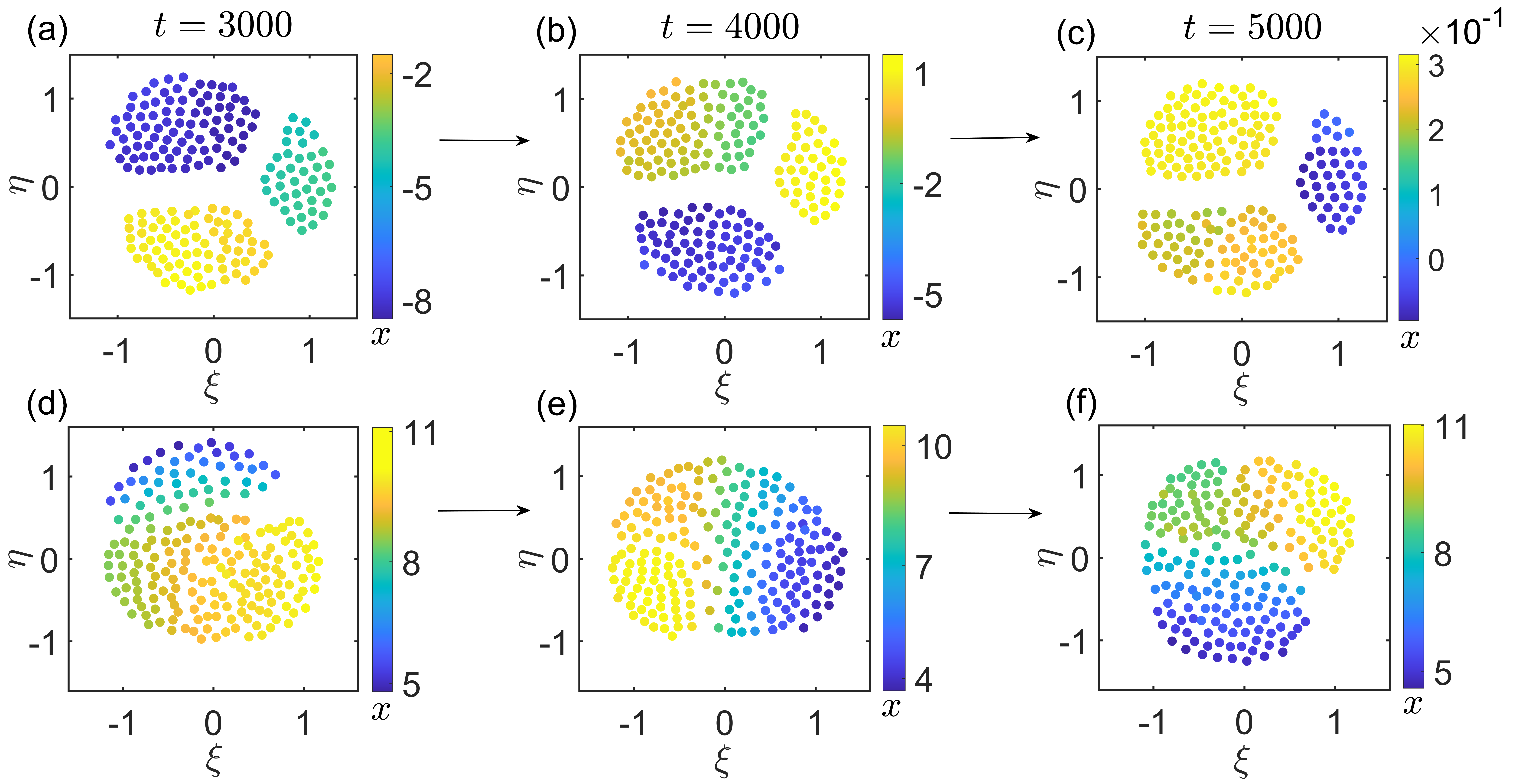}
    \caption{Splintered phase wave and phase wave states for swarmalators with chaotic R{\"o}ssler oscillator. Snapshots of the splintered phase wave state are shown at (a) $t=3000$, (b) $t=4000$, and (c) $t=5000$ time units for $K=0.012$. (d), (e) and (f) depict the snapshots of the phase wave state at $t=3000,\ 4000$, and $5000$ time units, respectively where $K=0.072$. In the entire simulation, we fix $J_1=2.0$, and $J_2=0.5$. We integrate Eqs.~\eqref{eq1}- \eqref{eq2} by Heun's method using $T=5000$ time units with step-size $dt=0.01$ for $N=200$ swarmalators.}
    \label{fig2}
\end{figure*}

To measure the amount of internal disorder among the swarmalators, we define the time average of the synchronization error of the R{\"o}ssler oscillator as  
\begin{equation}
\langle E \rangle= \left \langle  \sum_{i,j=1}^{N} \dfrac{| \mathbf{x}_j(t) - \mathbf{x}_i(t) |}{N(N-1)} \right \rangle_t, 
\label{eq5}
\end{equation}
where $\langle \cdots\rangle_t$ stands for the time average, after discarding the initial transients. The red curve (filled circles) in Fig.~\ref{fig1} refers to the time average synchronization error $\langle E \rangle$ as a function of the coupling strength $K$ delineated through the left $y$-axis. Our model reveals both stationary and non-stationary states. To distinguish them, we measure the mean velocity of the swarmalators as 
\begin{equation}
	\langle v\rangle = \Bigg \langle\frac{1}{N} \sum_{i=1}^{N} \sqrt{{{\Delta \xi_i}^2+{\Delta \eta_i}^2}} \Bigg \rangle_t,
	\label{eq6}
\end{equation}

where $(\Delta \xi_i, \Delta \eta_i)$ is the displacement of the $i$-th  swarmalator in the 2D plane in time interval $t$ to $t+dt$.
The blue curve (open circles) in Fig.~\ref{fig1} represents this quantity as a function of $K$ plotted in the log-linear scale in the right $y$-axis. With the understanding of these order parameters, we delve into examining the emerging collective states of our model.

\subsection{Emerging collective states}
We explore the synchronization and spatial properties of the swarmalator model. Based on numerical analysis, we find that the parameter $K$ plays a pivotal role in our case study, as the interaction strength among the R{\"o}ssler oscillators controls the synchronization phenomena, eventually determining the spatial pattern formation of the swarmalators. Consequently, we systematically adjust $K$ across a defined range, leading to the identification of five distinct long-term collective states: \textit{sync}, \textit{async}, \textit{splintered phase wave}, \textit{phase wave}, and \textit{chimera}. As the swarmalators diverge due to strong repulsion for $K<-0.13$ and remain synchronized for $K>0.3$, we meticulously focus on the region $K \in [-0.13,0.3]$ where we can objectify all the emerging states. Figures~\ref{fig1}(a)-(e) best illustrate these states where each of the swarmalators is represented by the scatter plots in the $(\xi,\eta)$ plane and their colors indicate the $x$ component of the R{\"o}ssler oscillator. We also scrutinize the pattern formation by coloring the swarmalators according to the other two components $y$ and $z$ and find similar outcomes (see Fig. 1 in the Supplementary Material \cite{supple} ).

\par \textbf{Async}: We start from the left panel in Fig.~\ref{fig1} where $K$ is negative. In the range $[-0.12,0.0]$, the swarmalators are moving (notice the velocity profile in Fig.~\ref{fig1} depicted by the blue curve) and they arrange themselves within a circular disc. Given that the interaction strength $K$ is negative, it hinders their ability to exhibit coherent behavior within the internal oscillations.
The corresponding internal dynamics remain desynchronized, exhibiting a wide range of $x$ values seen in  Fig.~\ref{fig1}(a). $\langle E \rangle$ is notably elevated in this region, suggesting a highly desynchronized behavior. We embellish this as the \textit{async}. Look at Movie 1 of the Supplementary Material \cite{supple} for the time evolution of the state.

\par
\begin{figure}
    \includegraphics[width=1.0\columnwidth]{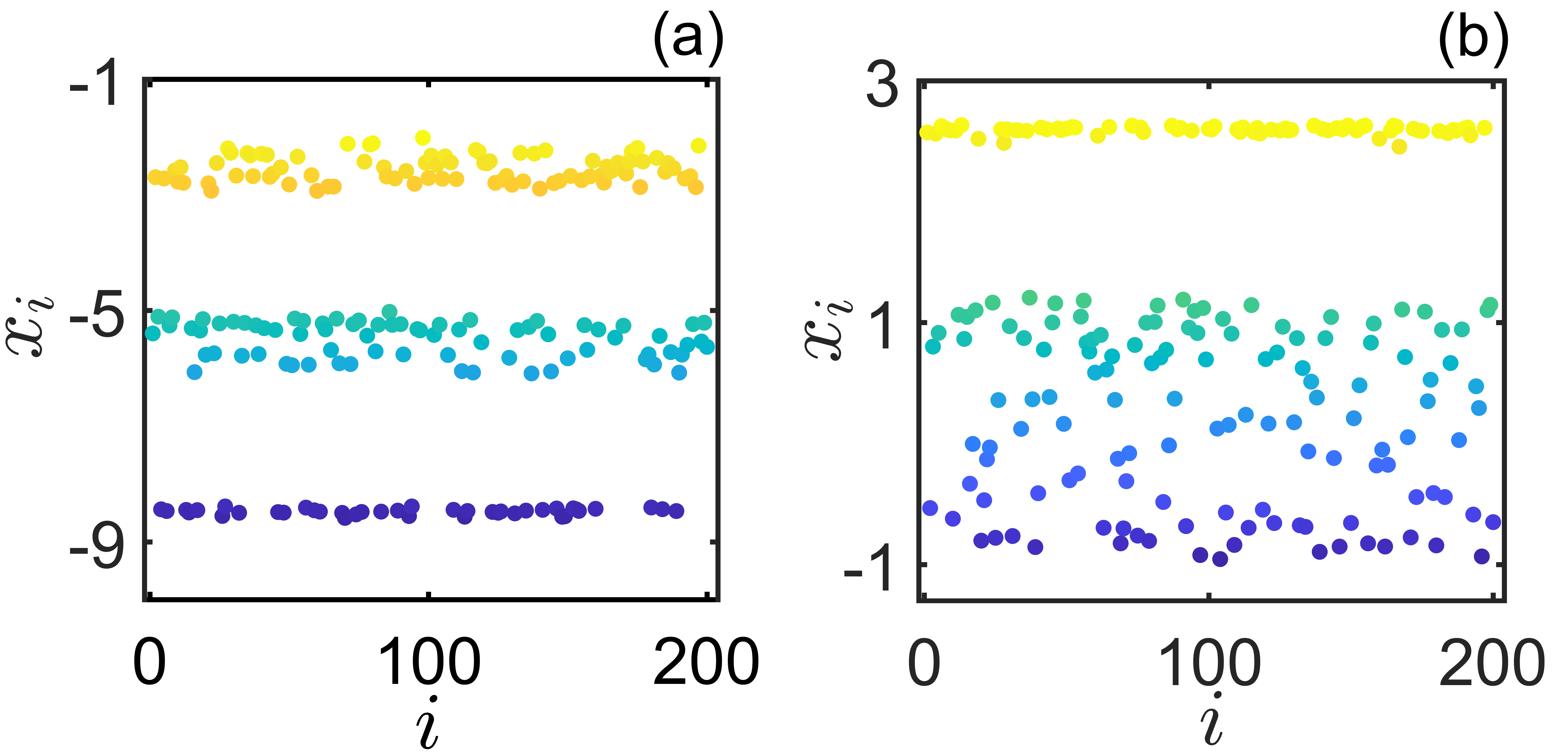}
    \caption{Scatter plots of the R{\"o}ssler $x$ components against their corresponding indices at $T=4000$ time units for $N=200$ swarmalators. (a) Splintered phase wave for $K=0.012$ where distinct clusters are observed. (b) Chimera for $K=0.072$ where two clusters are visible among which one is synchronized and in the other one swarmalators show asynchronous behavior. Heun's method has been used to integrate Eqs~\ref{eq1} and \ref{eq2} with step-size $dt=0.01$ to generate both the figures.} 
    \label{fig3}
\end{figure}

 \textbf{Splintered phase wave}: Moving to the right of Fig.~\ref{fig1}, as we increase the value of $K$ from 0, we note a discernible trend where the swarmalators form multiple clusters, each exhibiting motion confined within its respective cluster, and their activity never dies off. The region between the black and pink dotted vertical lines ($0.0<K<0.032$) illustrates the state specifically. If we recall the splintered phase wave of the phase-influenced swarmalator model \cite{o2017oscillators}, this state is the closest realization of that. For this, we refer to this state as \textit{splintered phase wave} (see Fig.~\ref{fig1}(b)). Figures~\ref{fig2}(a)-(c) display the snapshots of this state at different times which demonstrates the evolution over time. We further plot the $x$ components of the R{\"o}ssler oscillator as a function of the indices for this state in Fig.~\ref{fig3}(a). The clusters display distinct internal behaviors in this state, each isolated from the others. Find supplementary \cite{supple} Movie 2 for a better visualization of this state.

\par \textbf{Phase wave}: When we gradually increase $K$ from the splintered phase wave state, swarmalators change their positions in a very rapid manner. They are spatially attracted towards the ones having minimal synchronization error $E_{ij}$ and the disjoint cluster formation like the splintered phase wave, disappears. Compared to the other states, we observe that swarmalators move with a higher velocity in this state which we will discuss in detail in the subsequent sections. We refer to this state as the \textit{phase wave} (look into Fig.~\ref{fig1}(c)). They form a deformed disk-like structure, where the positions are distributed inside the disk and they are attracted to the ones having nearby internal dynamics (in the phase wave state reported with phase oscillators, they were distributed uniformly inside an annular ring). In Figs.~\ref{fig2}(d)-(f), we capture the snapshots at different times to analyze their movement. Also, see Movie 3 in Supplementary \cite{supple} for the time evolution of the phase wave state. We specify this region between the pink and the yellow dotted vertical lines ($0.032<K<0.116$) shown in Fig.~\ref{fig1} where the phase wave is realized.

\par \textbf{Chimera}: On further increment of $K$, we observe a bunch of swarmalators organize themselves within a cluster having similar internal dynamics (coherent $x$ values), and the rest of the swarmalators form another cluster with a wide range of $x$ values. All of them show feeble spatial movements within their specific clusters. (Find Movie 4 of the Supplementary Material \cite{supple}). The nature of this state is similar to the chimera state where coherent and incoherent behaviors coexist \cite{abrams2008solvable, abrams2004chimera, majhi2019chimera}. We represent this state as \textit{chimera state}. Figure~\ref{fig1}(d) illustrates this state by showing the snapshot in the $(\xi,\eta)$ plane. We observe this state when $0.116<K<0.176$. The region is highlighted by the yellow and the violet dotted vertical lines in Fig.~\ref{fig1}. For a clear picture, we plot the $x$ components of the R{\"o}ssler oscillators against their respective indices in Fig.~\ref{fig3}(b). The presence of one coherent group where the $x$ values are the same is observed along with an incoherent group where the $x$ values are distributed.

\par \textbf{Sync}: When $K$ is positively large ($K>0.176$), the R{\"o}ssler oscillators get synchronized and $\langle E \rangle$ goes to zero. The spatial movement of the swarmalators also diminishes, i.e., $\langle v\rangle=0$. On top of that, we observe all the swarmalators organize themselves inside a circular disk. They are synchronized at every instant, indicating that the collective internal dynamics become identical for all the swarmalators (look at Movie 5 in  Supplementary \cite{supple}). We mark this state as \textit{sync} which is represented by the scatter plots in Fig.~\ref{fig1}(e). 

Therefore, with increasing coupling strength $K$ the entire route from the async to the sync state can be depicted as \textbf{async $\rightarrow$ splintered phase wave $\rightarrow$ phase wave $\rightarrow$ chimera $\rightarrow$ sync}.

\subsection{Radii of the sync and async states}
\par \textbf {Sync}: Here, the agents do not exhibit any spatial movements. Moreover, each of them embarks on identical internal dynamics at every instant of time. One can capture this information regarding this state in terms of the velocity field, which contains all necessary details, i.e., their spatial positions and relative internal errors. The corresponding field can be illustrated with the following form $\mathbf{v}=\mathbf{v}(\mathbf{r},\Tilde{E}(\mathbf{r}))$. Since the spatial movement of the agents is absent, we can conclude the velocity field $\mathbf{v}=0$.

When examining the specifications of steady-state patterns, an insightful approach is involved in investigating the continuity equation. In this state, where the density $\rho (\mathbf{r},\tilde{E}(\mathbf{r}))$ represents the stationary distribution of swarmalators with positions $\mathbf{r}$ in $(\xi,\eta)$ plane and internal relative error $\tilde{E}(\mathbf{r})$, the divergence of the velocity field must be zero. This requirement is underscored by the normalizing property, stating that $\int \rho(\mathbf{r},\tilde{E}(\mathbf{r})) d\mathbf{r} =1$. Finally, we end up with a pair of simultaneous equations, 
\begin{align}
    \mathbf{v} &\equiv 0,\label{eq7} \\
    \nabla \cdot \mathbf{v} &\equiv 0. \label{eq8}
\end{align}
In Cartesian coordinates, the velocity field reads
\begin{align}\label{eq9}
&\mathbf{v}\left(\mathbf{r},\Tilde{E}(\mathbf{r})\right)=\int\Bigg(\left(\mathbf{\Tilde{r}}-\mathbf{r}\right)\left(A+J_1e^{-\Tilde{E}(\mathbf{\Tilde{r}})}\right)\nonumber\\
&-\dfrac{\mathbf{\Tilde{r}}-\mathbf{r}}{|\mathbf{\Tilde{r}}-\mathbf{r}|^{2}}\left(B-J_2 e^{-\Tilde{E}(\mathbf{\Tilde{r}})}\right)\Bigg)\rho\left(\mathbf{\Tilde{r}},\Tilde{E}(\mathbf{\Tilde{r}})\right)~d\mathbf{\Tilde{r}}.
\end{align}

In the sync state, the internal dynamics of the swarmalators become identical at every instant of time. When the relative internal error boils down, i.e., $\Tilde{E}\rightarrow 0$, the integral part of Eq.~\eqref{eq9} no longer depends on the relative error, rather it depends only on the spatial profile, which allows us to treat Eq.~\eqref{eq9} by excluding $\Tilde{E}$. Remembering this fact, Eq.~\eqref{eq9} can be rewritten as,
\begin{align}\label{eq10}              
&\mathbf{v}\left(\mathbf{r}\right)=\int\left(\left(\mathbf{\Tilde{r}}-\mathbf{r}\right)\left(A+J_1\right)-\dfrac{\mathbf{\Tilde{r}}-\mathbf{r}}{|\mathbf{\Tilde{r}}-\mathbf{r}|^{2}}\left(B-J_2 \right)\right)\nonumber\\
&~~~~~~~~~~~~~~~~~~~~~~~~~~~~~~~~~~~~~~~~~~~~~~~~~\rho\left(\mathbf{\Tilde{r}}\right)~d\mathbf{\Tilde{r}}.
\end{align}
Recalling Eq.~\eqref{eq8}, by taking the divergence of Eq.~\eqref{eq10} we get the following equality
\begin{align}\label{eq11}
&\nabla\cdot\mathbf{v}\left(\mathbf{r}\right)=\int \biggl(2(A+J_1)-2\pi\delta(\mathbf{\Tilde{r}}-\mathbf{r})(B-J_2)\biggr)\nonumber\\
&~~~~~~~~~~~~~~~~~~~~~~~~~~~~~~~~~~~~~~~~~~~~~~~~~\rho(\mathbf{\Tilde{r}})d\mathbf{\Tilde{r}}=0.
\end{align}
Simplifying Eq.~\eqref{eq11}, we obtain
\begin{equation}
    \int 2\pi \delta \left(\mathbf{r}-\mathbf{\Tilde{r}}\right)\rho(\mathbf{\Tilde{r}})d\mathbf{\Tilde{r}} =\dfrac{2(A+J_1)}{B-J_2}, \nonumber 
\end{equation}
which gives,
\begin{equation}\label{eq12}
    \rho(\mathbf{r}) = \dfrac{1}{\pi}\dfrac{A+J_1}{B-J_2}.
\end{equation}

\begin{figure}
    \includegraphics[width=\columnwidth]{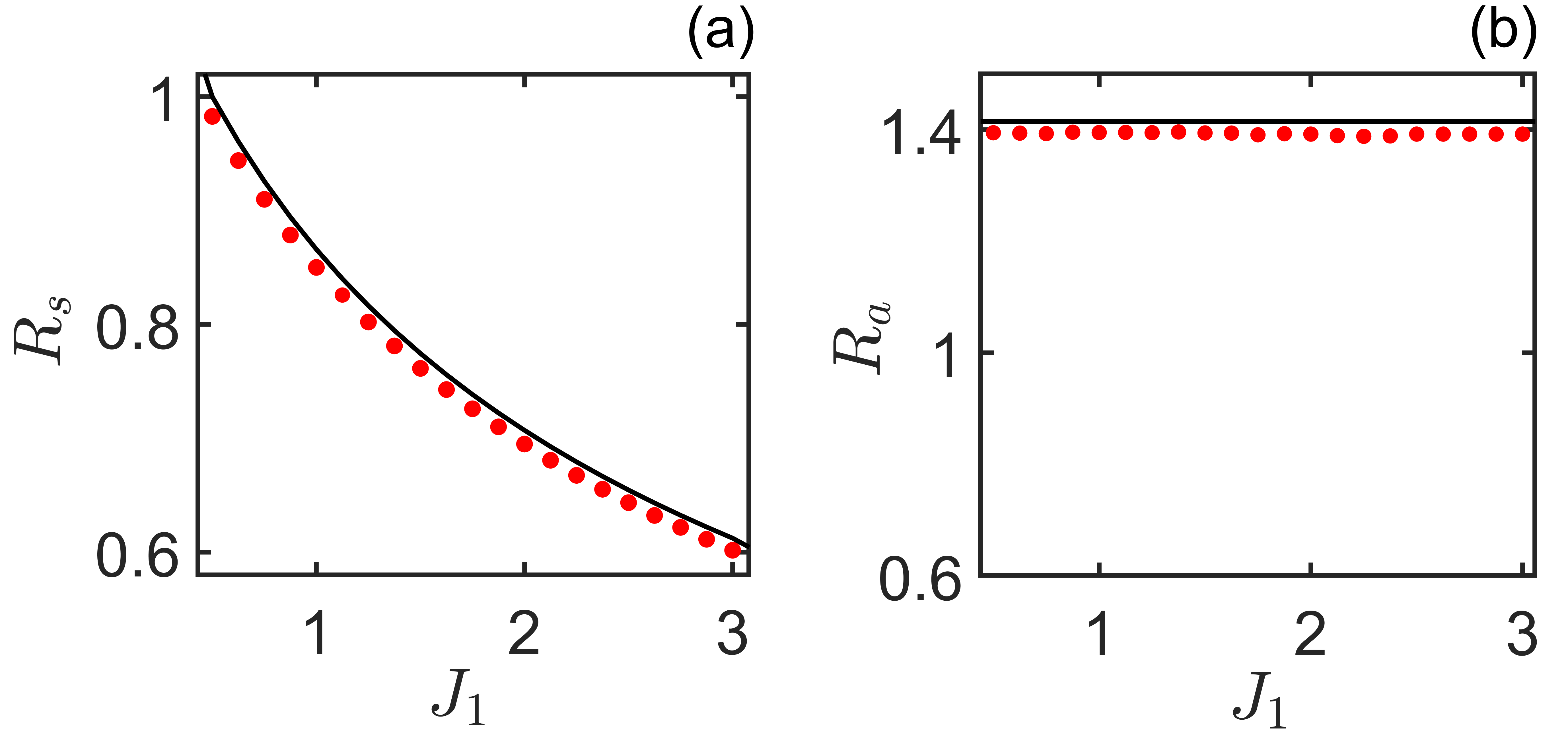}
    \caption{(a) Radius of the sync state as a function of $J_1$ for $K=0.25$. (b) The radius of the async state against $J_1$, where $K=-0.05$. The red dots correspond to the simulation results and the black curve represents the analytical measurement. Data were generated for $T=4000$ time units by integrating Eqs~\ref{eq1} and \ref{eq2} with Heun's method for $(dt, N)= (0.01,200)$.}  
	\label{fig4} 
\end{figure}

In this state, the swarmalators form a disc in the $(\xi,\eta)$ plane. The center of position is conserved by the symmetric pairs of changes of Eq.~\eqref{eq1}. So due to the rotational symmetry, the density of steady state can be written as $\rho_s\left(\mathbf{r}\right)=1/(\pi R_s^2)$ for $|\mathbf{r}|\leq R_s$. Comparing Eq.~\eqref{eq12} with this, we get the expression of the radius in the sync state as
\begin{equation}\label{eq13}
    R_s=\sqrt{\dfrac{B-J_2}{A+J_1}}.
\end{equation}
In Fig.~\ref{fig4}(a), we explore the dependence of the radius of the sync state on $J_1$. Furthermore, we collocate our analytical finding, as expressed in Eq.~\eqref{eq13}, with the numerical outcome, revealing a strong concurrence between them.

{\bf Async}: In this segment, we analyze the async state, where the relative error between the oscillators $(\Tilde{E})$ takes a tremendous nonzero positive value. As a result, the influence of $J_1$ and $J_2$ becomes insignificant, and therefore, their effects can be disregarded in this scenario. Hence, the term $e^{-\Tilde{{E}}}\rightarrow 0$ and the Eq.~\eqref{eq9} can be rewritten in the following form,
\begin{equation}\label{eq14}                    
\mathbf{v}\left(\mathbf{r}\right)=\int\Bigg(\left(\mathbf{\Tilde{r}}-\mathbf{r}\right)A-\dfrac{\mathbf{\Tilde{r}}-\mathbf{r}}{|\mathbf{\Tilde{r}}-\mathbf{r}|^{2}}B\Bigg)~\rho\left(\mathbf{\Tilde{r}}\right)~d\mathbf{\Tilde{r}}.
\end{equation}
Performing a comparable calculation with the sync state yields the form of the probability density function for the async state as follows,
\begin{equation}\label{eq15]}
    \rho(\mathbf{r})=\dfrac{1}{\pi}\dfrac{A}{B}.
\end{equation}
Therefore, the radius for the async state takes the form as,
\begin{equation}\label{eq16}
    R_a=\sqrt{\dfrac{B}{A}}.
\end{equation}
We validate our analytical finding given by Eq.~\eqref{eq16} with the numerical simulation in Fig.~\ref{fig4}(b).
This emphasizes that the radius of the async state is independent of $J_1$ (also $J_2$, from the expression in Eq.~\eqref{eq16}).

\begin{figure}[hpt]
    \includegraphics[width=1\columnwidth]{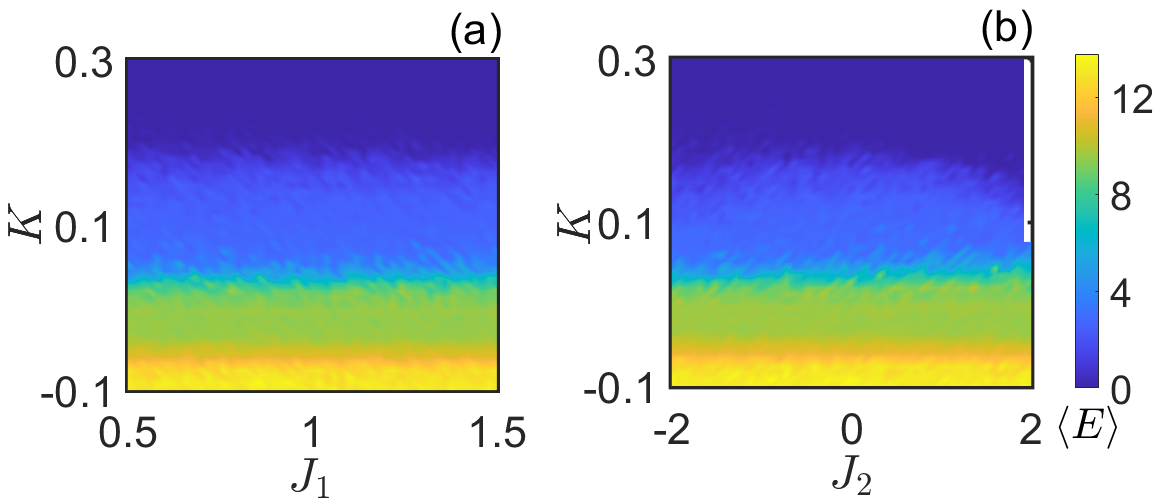}
    \caption{(a) Variation of $\langle E \rangle$ in the $J_1$-$K$ parameter space for $J_2=0.5$, and in (b) $J_2-K$ parameter space for $J_1=2.0$. The model is integrated with $N=200$ swarmalators using Heun's method for step-size $dt=0.01$ and $T=4000$ time units. After discarding the transients, the average synchronization error $\langle E \rangle$ is calculated with the last $25\%$ of data.}
    \label{fig5}
\end{figure}

\subsection{Dependence on system parameters}
We explore the impact of $J_1$ and $J_2$ on the dynamics of the swarmalators. For this, we choose to vary one of $J_1$ and $J_2$ with the coupling strength $K$, keeping the other fixed. In Fig.~\ref{fig5}(a), simultaneously varying $J_1$ with $K$, while fixing $J_2$ at 0.5, we observe the system's dynamical behavior. In Fig.~\ref{fig5}(b), we vary $J_2$ over a range of $K$ by keeping $J_1=2.0$. The color bar signifies the time average synchronization error $\langle E \rangle$ of the R{\"o}ssler oscillator. From the $J_1-K$ parameter space, it is evident that the collective behavior of the system is independent of $J_1$ and primarily depends on $K$. From Fig.~\ref{fig5}(b), we observe a similar scenario except in the upper right corner, where the system becomes highly unstable and collapses for a high $J_2$ value. It is worth reminding ourselves again that the spatial attraction and repulsion terms are given by $\left(A+J_1 e^{-E_{ij}}\right)$ and $\left(B-J_2 e^{-E_{ij}}\right) / |\mathbf{r}_j-\mathbf{r}_i |^{2}$, respectively. When the value of $K$ is large enough to attain synchrony among the R{\"o}ssler oscillators, the repulsion term eventually becomes $B-J_2$. We should have $J_2<B$, or else the repulsive term acts like an attraction term. In that case, some swarmalators might collide with each other which is not feasible for a swarmalator system. Since we have fixed $B=2$ in our work, we see that when $K$ is sufficiently large and $J_2\rightarrow 2$, there is an unbounded region (the white region in Fig.~\ref{fig5}(b)).

\subsection{Velocity profiles of the states}
\begin{figure}
    \centering
    \includegraphics[scale=0.22]{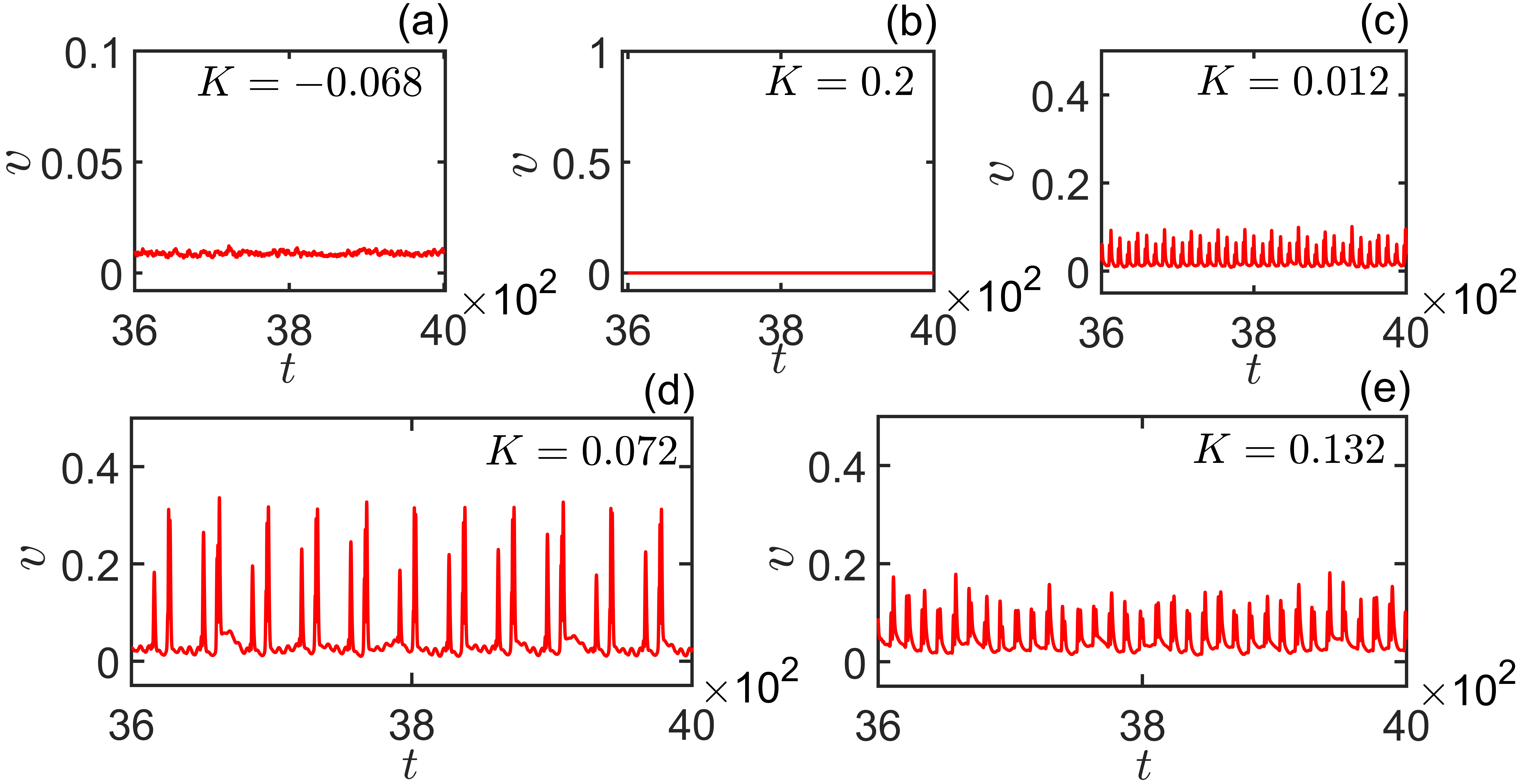}
    \caption{The temporal evolution of the velocity of the center of positions $(v)$ regarding the emerging states. (a) Async $(K=-0.068)$, (b) sync $(K=0.2)$, (c) splintered phase wave $(K=0.012)$ , (d) phase wave $(K=0.072)$, and (e) chimera $(K=0.132)$. The model is integrated with $N=200$ swarmalators using Heun's method with step-size $dt=0.01$ for $T=4000$ time units. We show the velocity time series for the last $10\%$ data for clarity.}
    \label{fig6}
\end{figure}

In the preceding sections, we studied several emerging patterns exhibited by our model and thoroughly examined their dynamics. The most remarkable outcome is the discovery of numerous states that have already been reported in the phase-influenced swarmalator model. In addition to the amplitude dynamics of the underlying R{\"o}ssler oscillator, we notice the swarmalators exhibiting various spatial movements in the emerging states. One can explore and distinguish these profiles in terms of the velocity of the center of positions $(v)$ for each state. Previously in Fig.~\ref{fig1}, we have captured the essence of nonzero time-averaged velocity over a range of coupling strength $K$. We showcase the corresponding temporal evolution of the ensemble displacement for each state in Fig.~\ref{fig6}. The swarmalators show very small displacement over time in the async state as observed in Fig.~\ref{fig6}(a), whereas, in the sync state, swarmalators attain steady positional configuration, i.e., $v= 0$. Look at Fig.~\ref{fig6}(b) for the steady behavior of $v$ in the sync state for $K=0.2$. In the remaining states, the overall structure pulsates over time. Notice the spiking behavior in the velocity profiles for these states (see Figs.~\ref{fig6}(c)-(e)). When the swarmalators form the phase wave, they frequently undergo large compression and expansion. Swarmalators try to minimize the difference between their internal states when $K$ is positive. They get spatially attracted towards each other when this difference decreases. On the other hand, when they come too close to each other, the spatial repulsion function disperses them away which gives a sudden jump in the velocity and the loop continues. This gives rise to some kind of positional instability in the system. We observe the spiking tails are significantly higher in the phase wave (Fig.~\ref{fig6}(d)) compared to both the splintered phase wave (Fig.~\ref{fig6}(c)) and chimera state (Fig.~\ref{fig6}(e)).

\begin{figure*}
    \centering
    \includegraphics[width=2\columnwidth]{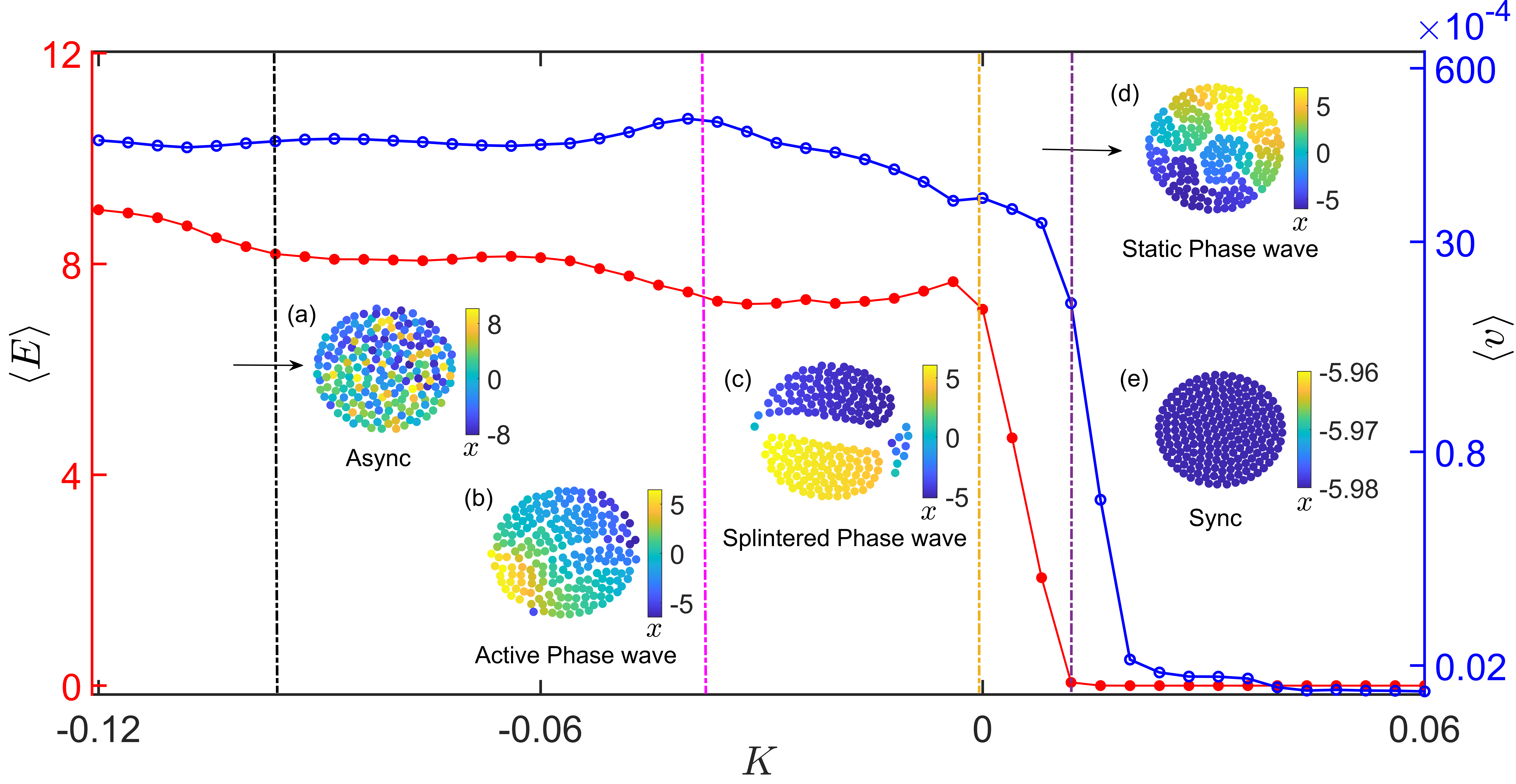}
	\caption{Variation of average synchronization error $\langle E \rangle$ (filled red circle) of period one R{\"o}ssler oscillator and the mean positional velocity of the swarmalators $\langle v \rangle$ (open blue circle) with the coupling strength $K$. We choose the parameter values as $a=b=0.1,\ c=4.0$ and fix $J_1=2.0,\ J_2=0.5$, $A=1.0$ and $B=2.0$. Dotted vertical lines separate the regimes between different emerging states. Snapshots are shown in (a) async for $K=-0.12$, (b) active phase wave for $K=-0.05$, (c) splintered phase wave for $K=-0.02$, (d) static phase wave for $K=0.0$, and (e) sync for $K=0.04$. The color bar signifies the $x$ component of the R{\"o}ssler oscillator. We integrate Eq.~\eqref{eq1} and Eq.~\eqref{eq2} by Heun's method for $(dt,T, N)=(0.01,4000,200)$. In all cases, swarmalators are initially placed inside a box with dimension $[-1,1]\times [-1,1]$ uniformly at random. All the resultant states are observed after a long transient. $\langle E \rangle$ and $\langle v \rangle$ have been calculated with the last $25 \%$ data after discarding the transients with $10$ realizations of initial conditions.} 
	\label{fig7} 
\end{figure*}

\section{Analysis on periodic oscillation}
\label{section4}
After investigating the swarmalators embedded with chaotic dynamics of the R{\"o}ssler oscillator, we explore them with the period one behavior. For this, we fix the parameters as $a=0.1$, $b=0.1$, and $c=4.0$ where the R{\"o}ssler oscillator shows one periodic behavior. 

Similar to the chaotic region, here also, we consistently adjust the oscillators' interaction strength $K$ over a finite range and observe the variation of average synchronization error $\langle E \rangle$ and the mean velocity $\langle v \rangle$ of the swarmalators to distinguish different emerging collective behaviors. In Fig.~\ref{fig7}, we demonstrate the variation of $\langle E \rangle$ and $\langle v \rangle$ by filled red circles and open blue circles, respectively. We again observe the async and sync states for extreme values of $K$. Both the async and sync states (see Figs.~\ref{fig7}(a) and (e)) resonate similarly to the case of the chaotic oscillator. The async state is found when $-0.12<K<-0.096$. What is novel about the periodic region is the discovery of three types of phase wave states, akin to those observed in the phase-influenced swarmalator model. Inside the region highlighted by the black and pink dotted vertical lines in Fig.~\ref{fig7} (where $-0.096<K<-0.038$), we observe the swarmalators coordinating themselves within a disk and each one of them tries to form clusters with another having coherent internal dynamics, characterized by the nearest $x$ values. As a whole, they follow a circular-like movement in the $(\xi,\eta)$ plane. Fig.~\ref{fig7}(b) depicts this state. We denote this state as the {\it active phase wave} due to the continuous movements of swarmalators from one portion to another inside the disk. When we increase $K$ further from $-0.038$, we capture the swarmalators to form disjoint clusters and move in the $(\xi, \eta)$ plane inside their respective groups. We identify this state as the splintered phase wave (notice Fig.\ref{fig7}(c)) as before. The extent of this state is delineated by pink and yellow dotted vertical lines ($-0.038<K<0.0$). The difference between the active and the splintered phase wave is best understood through Movie 6 of the Supplementary Material \cite{supple}. The first stationary state ($\langle v \rangle \approx 0$) occurs for the case when $K=0$. The swarmalators become static by their position within the regions highlighted by the yellow and violet dotted vertical lines where $0.0<K<0.012$. They arrange themselves into small groups where they are situated nearby to the ones having minimal relative error. We refer to this state as {\it static phase wave} which is depicted by Fig.~\ref{fig7}(d). Finally, we achieve synchrony at $K=0.012$, after which all the swarmalators become completely static by their positions ($\langle v \rangle = 0$) and resonate with similar internal dynamics at an instant of time. 
Therefore the overall dynamical route in this case becomes \textbf{async $\rightarrow$ active phase wave $\rightarrow$ splintered phase wave $\rightarrow$ static phase wave $\rightarrow$ sync.} 

\begin{figure}
    \includegraphics[width=1.02\columnwidth]{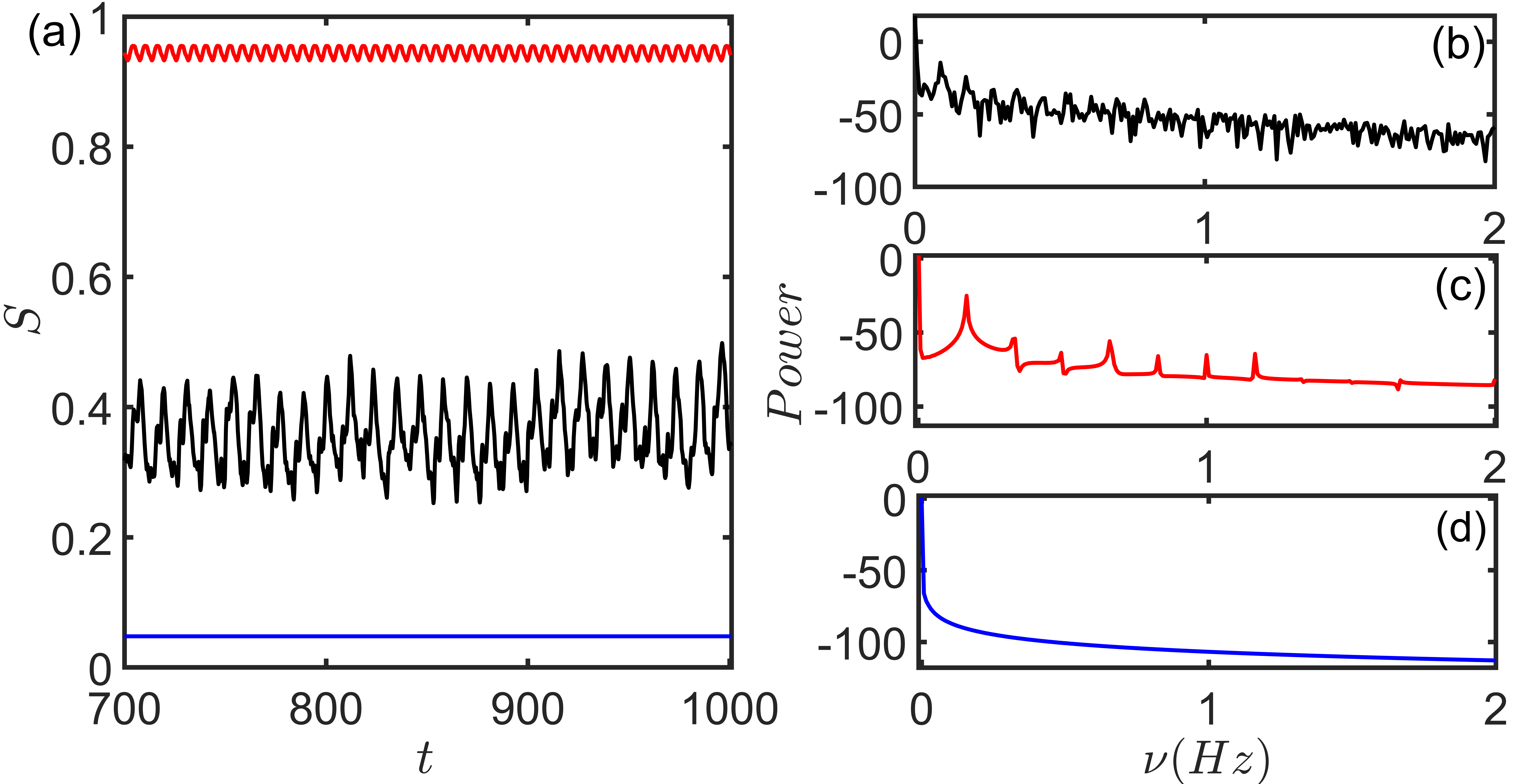}
    \caption{(a) Time series of $S$ for $K=-0.1$ (black), $K=0.0$ (red), and $K=0.1$ (blue). (b), (c) and (d) are the normalized power spectrum for $K=-0.1,0$, and $0.1$, respectively. We take $N=200$ swarmalators and run simulations for $T=4000$ time units with step-size $dt=0.01$ using Heun's method.} 
    \label{fig8}
\end{figure}

To measure the correlation between swarmalators' spatial angle $\phi_j=\tan^{-1}(\eta_j/\xi_j)$ and the oscillators' phase $\theta_j=\tan^{-1}(y_j/x_j)$, we define the following order parameters as:
\begin{equation}
    S_\pm e^{i \psi \pm} = \frac{1}{N} \sum_{j=1}^{N} e^{i (\theta_j \pm \phi_j)},
    \label{eq17}
\end{equation}
which quantifies the correlation between the spatial and internal dynamics of the swarmalators. We take the maximum of $S_{\pm}$ and define $S=\max\{S_+,S_-\}$. A non-zero value of $S$ signifies the presence of a correlation between the swarmalators' spatial position to the oscillators' internal dynamics.
First, we observe the time evolution of $S$ at different values of $K$. For the $K<0$ region, we notice the chaotic nature of $S$ depicted by the black curve in Fig.~\ref{fig8}(a). In the positive $K$ region, the oscillation of $S$ completely dies and does not vary with time (see the blue line in Fig.~\ref{fig8}(a)). For $K=0$, when the dynamics of the R{\"o}ssler oscillators do not affect the spatial positions, we observe the periodic nature of $S$ with time (look at the red curve in Fig.~\ref{fig8}(a)). We have also looked at the power spectrum to validate the chaotic (Fig.~\ref{fig8}(b)), periodic (Fig.~\ref{fig8}(c)), and constant (Fig.~\ref{fig8}(d)) behaviors of $S$ for $K=-0.1,0$, and $0.1$, respectively.

\section{Discussions}
\label{section5}
It is worth mentioning that till now, swarmalators have been studied through the vision of phase oscillators. The combined effect of phase-dependent spatial aggregation and position-dependent phase synchronization defines a novel phenomenon and leads to various rich spatio-temporal patterns. Some of these stationary and non-stationary patterns can be found in many real-world systems like Japanese tree frogs \cite{aihara2014spatio}, magnetic domain walls \cite{hrabec2018velocity}, Janus matchsticks \cite{chaudhary2014reconfigurable}, robotic swarms \cite{esam2021ieee, barcis2020sandsbots} etc. Nowadays, researchers investigate this system by applying suitable interacting functions, coupling schemes, network structures, etc. Our work's novelty lies in exploring the system from the perspective of the amplitude oscillators. 
\par In this article, we have studied the trade-off between the spatial dynamics over the chaotic (and periodic) internal dynamics of the swarmalators and vice-versa. We have modeled swarmalators so that their interactions are all-to-all, and their internal dynamics are governed by the R{\"o}ssler system. We here claim that not only the phase oscillator, we can visualize the swarmalator field from the perspective of the amplitude oscillator also. We have encountered the presence of most of the states akin to those observed in the phase-inspired swarmalator model~\cite{o2017oscillators}. Our model comprises three parameters $K$, $J_1$, and $J_2$ which mainly regulate the long-term behavior of the swarmalator system. We analytically derived the radius of the sync and async states which confirms our numerical findings and emphasize the originality of our work. We are unable to solve the radius of the other states as their structure continuously changes with time. Subsequently, we explore the system within the non-chaotic domain (period one), leading to the emergence of five distinct long-term collective states.  
\par In summary, our model aims to provide a fresh outlook on swarmalator dynamics. One can think about how these internal dynamics can be further modified using the higher-order coupling scheme for various parameter values. We expect there might be other possible emerging states. Also, we anticipate that using a hyperchaotic system (characterized by more than one positive Lyapunov exponents) as swarmalators' internal dynamics, or complex switching mechanisms between multistable states, our system might exhibit a rich variety of complex behaviors and diverse emerging states, surpassing the limitations associated with a basic chaotic system. Implementing the vision range of each entity will open a new avenue for study on the swarmalator. On top of that, we can implement the angular dynamics under the influence of inertia on the bare swarmalator system\cite{o2017oscillators}. A pertinent future goal to explore would involve examining alternative physically feasible structures for both spatial and internal interactions. The analytical feasibility of our research offers a platform for investigating these avenues in the coming times.

\begin{figure*}[hpt!]
	\centering
	\includegraphics[width=2\columnwidth]{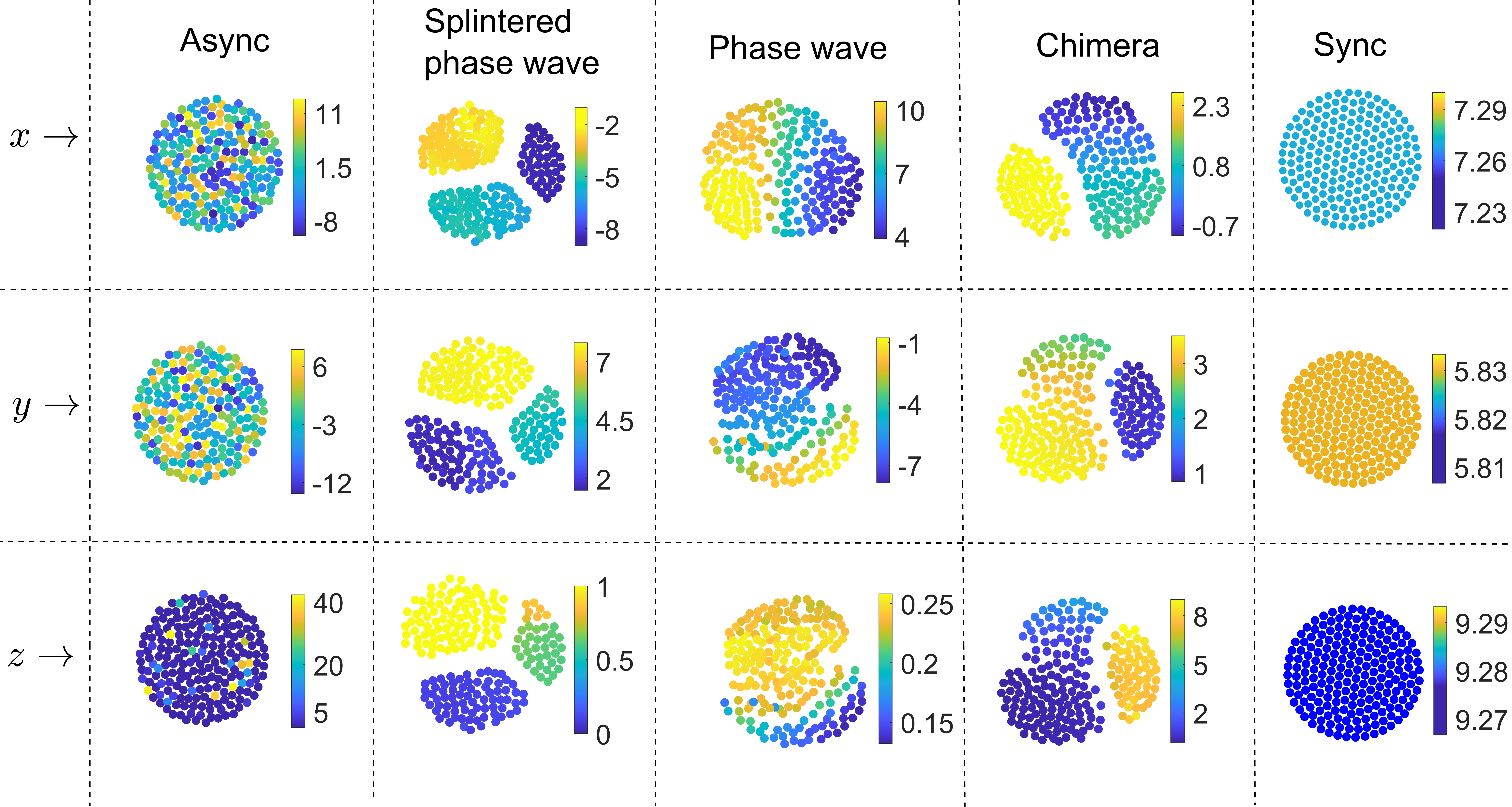}
	\caption{Temporal evolution of the five long-term emerging states for chaotic R{\"o}ssler attractor $(a=b=0.2,~c=5.7)$. The states in each row are depicted concerning the $x$, $y$, and $z$ components of the R{\"o}ssler attractor, represented by the color bar. The model is integrated by Heun's method with $(dt, T, N)=(0.01,4000,200)$.}
	\label{fig9}
\end{figure*}
\section*{Supplementary materials}
\section*{Movies}

We explore the swarmalator system from a new perspective. We investigate the two-dimensional swarmalator system through the lens of amplitude-mediated oscillatory dynamics. We keep $K$ as the free parameter and vary across a wide range to observe different emerging patterns and their behavior. To best visualize the states, we make movies of them to show how they evolve with time. To do this, we integrate the governing equations by Heun's method and fix $(dt, T, N)=(0.01, 4000, 200)$. Initially, the swarmalators are placed inside a square of dimension $[-1,1]\times[-1,1]$, uniformly at random and their internal dynamics are governed by the $x$ component of the R{\"o}ssler attractor. \\ 

In all the movies, we fix $J_1=2.0$ and $J_2=0.5$. From Movie 1 to Movie 5, we show the time evolution of the states for which we choose the R{\"o}ssler parameters from the chaotic region such as $a=b=0.2$ and $c=5.7$. \\

In Movie 1, we show the async state where the swarmalators move and their internal dynamics remain desynchronized. They form a disc and are uniformly distributed in the $(\xi, \eta)$ plane.\\

Movie 2 illustrates the splintered phase wave, where we observe a noticeable pattern whereby the swarmalators assemble into numerous clusters, each displaying motion confined within its respective cluster, and their activity persists indefinitely.\\

In Movie 3, we demonstrate the time evolution of the phase wave state, where the swarmalators move with a high velocity and are spatially attracted towards the one having minimal synchronization error $E_{ij}$.\\

Movie 4 depicts the chimera state beautifully, where coherence and incoherence clusters coexist. The coherence is determined by analogous internal dynamics among the swarmalators.\\

In Movie 5, we capture the time evolution of the sync state, i.e., the swarmalators become static by their position, and their internal behavior becomes identical at every instant.\\

The parameter values used are listed below:
\begin{itemize}
	\item Async: $K_3=-0.068$\
	\item Splinterded phase wave: $K_3=0.012$\
	\item Phase wave: $K_3=0.072$\
	\item Chimera: $K_3=0.132$\
	\item Sync: $K_3=0.2$\
\end{itemize}

Movie 6 portrays the difference between the active phase wave and the splintered phase wave state, which we find in the period-1 region of the R{\"o}ssler system. We fix the R{\"o}ssler system's parameters at $a=b=0.1$ and $c=4.0$.
\par Please see the online version in journal page to see the movies.
\section*{Collective five states with respect to $x$, $y$ and $z$ variables}
As stated in the manuscript, we utilized the $x$ component of the R{\"o}ssler attractor to color the swarmalators, and we observed that similar outcomes arise when using the $y$ and $z$ components. In Fig.~\ref{fig9}, we have shown the time evolution of the five emerging states by coloring the swarmalators according to $x$ (upper row), $y$ (middle row), and $z$ (bottom row) components of the chaotic R{\"o}ssler attractor.

\bibliographystyle{apsrev4-2}
\providecommand{\noopsort}[1]{}\providecommand{\singleletter}[1]{#1}%

\end{document}